\RequirePackage{fix-cm}
\documentclass[smallextended]{svjour3}       
\smartqed  
\usepackage{hyperref}
\usepackage{array} 
\usepackage{xcolor}
\usepackage{listings}
\usepackage[most]{tcolorbox}
\usepackage{xspace}
\usepackage[authoryear]{natbib}              %
\usepackage{graphicx}
\usepackage{xcolor}
\usepackage{ifthen}
\usepackage{rotating}
\usepackage{booktabs}
\usepackage{amsmath}
\usepackage{multirow}

\newcommand{\mysubsection}[1]{\vspace{1.25mm}\noindent\textit{\textbf{#1.}}}
\newcommand{\mysubsubsection}[1]{\vspace{1.25mm}\noindent\textit{#1.}}
\newcommand{\smalltt}[1]{{\small \texttt{#1}}}

\begin{document}
\title{Following Breadcrumbs in Code}
\subtitle{What Accidentally Committed Ad-Hoc Logs Reveal about Developer Comprehension}


\author{Yi-Hung Chou \and Boyuan Jiang \and Vidit Jain \and Yiyang Min \and April Yi Wang \and James A. Jones}


\institute{
Y. Chou\textsuperscript{1} \and 
B. Jiang\textsuperscript{1} \and 
V. Jain\textsuperscript{1} \and 
Y. Min\textsuperscript{2} \and 
A. Wang\textsuperscript{3} \and
J. A. Jones\textsuperscript{1} \\
\textsuperscript{1}University of California, Irvine, USA \\
\textsuperscript{2}Amazon, Toronto, Canada \\
\textsuperscript{3}ETH Zürich, Switzerland \\
\email{\{yihungc1, boyuanj, viditj3, jajones\}@uci.edu, rogermyy@amazon.com, april.wang@inf.ethz.ch}
}
\date{Received: date / Accepted: date}

\maketitle

\begin{abstract}
Developers frequently insert temporary print or log statements—ad-hoc logs—to better understand program behavior at runtime, particularly when facing unexpected issues or complex control flows. Despite being a nearly universal practice, systematic study has been limited because these logs are ephemeral: they usually remain only in local environments and are removed before code is committed, making them difficult to capture. In this work, we addressed this challenge by mining accidental commits where developers unintentionally left ad-hoc logs and later deleted them, and by analyzing live-streamed programming sessions to observe their use in practice. 
Using these methods, we constructed a large dataset across three major programming languages (Java, JavaScript, and Python), enabling the large-scale investigation of logging practices. 
Our analysis reveals both common and language-specific patterns in where and how developers rely on ad-hoc logs. Across languages, ad-hoc logs tend to appear in program regions that are harder to reason about at runtime. We also identify distinctive language-level patterns, such as frequent use in asynchronous and callback functions in JavaScript and in thread-related classes in Java. In addition, functions containing ad-hoc logs generally have higher cyclomatic complexity than the overall function population. Production logs show a similar association, consistent with logging serving as a means of observing runtime behavior in structurally complex functions. Together, these findings provide empirical insight into developers’ runtime comprehension practices and offer a valuable dataset for researchers and tool builders seeking to better support debugging and logging.
\end{abstract}

\section{Introduction\label{sec:introduction}}
Logging statements in software development generally serve two distinct purposes. (1) The first concerns statements that remain in the codebase after deployment, commonly referred to as \textbf{production logging statements}. 
These persistent statements are used to monitor run-time behavior and provide diagnostic information, thereby supporting developers in maintaining and troubleshooting software in production.  (2) The second, and arguably less recognized yet commonly seen, use of printf statements involves their temporary insertion during development. In this context, developers rely on them as lightweight tools for debugging, testing, or understanding program behavior. Such statements, or \textbf{ad-hoc logs}, are typically removed prior to deployment, and this practice is often referred to as \textbf{printf debugging}.

Logging statements, particularly those that are meant to survive into production, have been extensively studied in the software engineering community, as shown in the survey conducted by \cite{gholamian_comprehensive_2022} and ~\cite{chen_survey_2022}.
For example, \cite{fu_where_2014} investigated where developers place their production logs for monitoring purposes, while \cite{li_are_2023} conducted manual inspections and surveys to evaluate the readability of logging statements. Building on such motivations, researchers have also proposed methods and tools to automate the improvement of production logging practices. For instance, \cite{zhu_learning_2015} developed tools to support developers in making informed decisions when inserting production logs.

   
However, to the best of our knowledge, the second type of logs—ad-hoc logs—remains largely understudied in the software engineering literature. 
Prior observational studies report that developers use ad-hoc logs as one of the most common strategies for program comprehension and debugging~\citep{alaboudi_exploratory_2019, alaboudi_what_2023, jiang_log-it_2023, beller_dichotomy_2018, siegmund_studying_2014}. Despite their recognized importance, the transient and local nature of these ad-hoc logs has constrained the community’s ability to systematically study and understand them.

Although the collection of transient and localized ad-hoc log statements may initially appear infeasible, we argue that two inherent properties of such logs enable their systematic and large-scale identification. First, ad-hoc logs are, by design, not intended to remain in production software. Nevertheless, developers are not immune to errors, and prior empirical studies have shown that inadvertent commits of unintended code artifacts are common in practice~\cite{wen_empirical_2020,meli_how_2019}. As a result, when ad-hoc logs are mistakenly introduced into a repository, developers typically issue subsequent corrective commits to remove these statements, thereby leaving recoverable traces. Second, qualitative evidence reported in~\cite{chou_learning_2025} indicates that ad-hoc logs are most often implemented through lightweight, native output functions (e.g., \textit{System.out.println}) rather than standardized and comparatively heavyweight logging frameworks such as \textit{log4j}. This implementation choice not only distinguishes ad-hoc logs from production-level logging but also makes their detection more tractable. By capitalizing on these properties, our research leverages the occurrence of ad-hoc logs within corrective commits as a serendipitous opportunity to address a critical research gap: the large-scale collection and analysis of ad-hoc log statements as a means to infer developers’ comprehension processes and debugging challenges during software development.


\begingroup

\definecolor{adhocdeletedbg}{RGB}{245,235,225}
\definecolor{adhockw}{RGB}{180,95,0}
\definecolor{adhocstr}{RGB}{120,60,140}
\definecolor{adhoccom}{RGB}{100,100,100}
\definecolor{adhocframe}{RGB}{210,210,210}
\definecolor{adhocgutter}{RGB}{90,90,90}

\lstdefinestyle{adhocbase}{
  basicstyle=\ttfamily\small,
  showstringspaces=false,
  keepspaces=true,
  columns=fullflexible,
  frame=none,
  aboveskip=0pt,
  belowskip=0pt,
  xleftmargin=0pt
}

\lstdefinestyle{adhocpython}{
  style=adhocbase,
  language=Python,
  keywordstyle=\color{adhockw},
  stringstyle=\color{adhocstr},
  commentstyle=\color{adhoccom}
}

\lstdefinestyle{adhocjava}{
  style=adhocbase,
  language=Java,
  keywordstyle=\color{adhockw},
  stringstyle=\color{adhocstr},
  commentstyle=\color{adhoccom}
}

\lstdefinestyle{adhocjs}{
  style=adhocbase,
  language=Java,
  morekeywords={console,log},
  keywordstyle=\color{adhockw},
  stringstyle=\color{adhocstr},
  commentstyle=\color{adhoccom}
}

\newenvironment{adhoccodebox}[1]
{
  \begin{tcolorbox}[
    enhanced,
    colback=white,
    colframe=adhocframe,
    boxrule=0.4pt,
    arc=4pt,
    left=10pt,
    right=10pt,
    top=10pt,
    bottom=10pt
  ]
  \lstset{style=#1}
}
{
  \end{tcolorbox}
}

\newlength{\adhocgutterwidth}
\setlength{\adhocgutterwidth}{1.4em}
\newlength{\adhocgapwidth}
\setlength{\adhocgapwidth}{0.35em}

\newcommand{\adhocdel}[2]{%
  \par\noindent
  \colorbox{adhocdeletedbg}{%
    \makebox[\dimexpr\linewidth-2\fboxsep\relax][l]{%
      \rule{0pt}{1.9ex}%
      \makebox[\adhocgutterwidth][c]{\color{adhocgutter}\sffamily\bfseries\footnotesize -}%
      \hspace{\adhocgapwidth}%
      {\ttfamily\small #1#2}%
    }%
  }%
  \par
}

\begin{figure}[t]
\centering

\begin{minipage}{0.95\linewidth}
\textbf{Python: lenstronomy/lenstronomy}\\
\textbf{Message:} removing debugging print statements

\vspace{0.35em}
\begin{adhoccodebox}{adhocpython}
\begin{lstlisting}
if check_convention and not self.ignore_observed_positions:
    kwargs_lens = self._convention(kwargs_lens)
\end{lstlisting}
\adhocdel{\hspace*{-0.3em}}{print(\textquotesingle{}I am leaving ray shooting partial\textquotesingle{})}
\end{adhoccodebox}
\end{minipage}

\vspace{1em}

\begin{minipage}{0.95\linewidth}
\textbf{Java: liquibase/liquibase}\\
\textbf{Message:} DAT-3587 Remove println

\vspace{0.35em}
\begin{adhoccodebox}{adhocjava}
\begin{lstlisting}
protected void doMigration() throws Exception {

    if (COMMANDS.HELP.equalsIgnoreCase(command)) {
\end{lstlisting}
\adhocdel{\hspace*{1.5em}}{System.out.println("HERE!!!!!!!!!!!!!!!!!!");}
\begin{lstlisting}
        printHelp(System.err);
\end{lstlisting}
\end{adhoccodebox}
\end{minipage}

\vspace{1em}

\begin{minipage}{0.95\linewidth}
\textbf{JavaScript: ellismarkf/lark}\\
\textbf{Message:} remove console logs

\vspace{0.35em}
\begin{adhoccodebox}{adhocjs}
\begin{lstlisting}
option.destinationArrivalHour = destinationHour;
option.destinationArrivalMins = destinationMins;
\end{lstlisting}
\adhocdel{}{console.log(\textquotesingle{}is this code running?\textquotesingle{});}
\end{adhoccodebox}
\end{minipage}

\caption{Examples of accidentally committed ad-hoc logs collected in our dataset.}
\label{fig:adhoc-log-examples}

\end{figure}

\endgroup

This study sheds light on the subject by examining accidental commits in version-control systems like GitHub, where developers explicitly remove ad-hoc logs, as noted in commit messages (e.g., “remove console.log” or “delete print”). Figure~\ref{fig:adhoc-log-examples} provides examples of ad-hoc logs and the corresponding commit messages. To gain deeper insight into how these logs are used, we also conducted a qualitative analysis of 36 hours of developers’ live-stream sessions collected by \cite{alaboudi_exploratory_2019}. Unlike the GitHub commits, which only capture mistakenly preserved logs, the live-stream data reveals all instances of temporary logging and their practical use. This comparison helps us better understand the situations and comprehension challenges in which developers rely on these logs. In short, the GitHub data offers broad coverage and scalability, while the live-stream data provides depth and nuanced qualitative understanding.

Our analysis uncovered several noteworthy patterns across three programming languages—Java, JavaScript, and Python. One recurring practice was that developers frequently structured their console outputs in ways that allowed them to quickly identify and compare program states or execution flows in real time. We also observed that ad-hoc logging was disproportionately concentrated in certain types of code locations. For instance, in JavaScript, developers inserted print statements into asynchronous and callback functions more often than would be expected given the overall distribution of such functions in the dataset, suggesting that these contexts demanded additional logging to clarify their behavior. Similarly, in Java, developers placed print statements into thread-related classes more than the overall distribution of such classes in the sample dataset. Furthermore, across three programming languages, ad-hoc logs were used more frequently in functions with high Cyclomatic Complexity than in simpler ones, highlighting their role in aiding comprehension of challenging code segments.

Our study makes the following contributions through extending work in \cite{chou_learning_2025}: 
(1) We extended the empirical study for understanding ad-hoc logs across three languages, guided by research questions informed by prior literature.
(2) We analyzed developers' live-coding sessions from ~\cite{alaboudi_exploratory_2019, alaboudi_what_2023} 
to gain insights into their ad-hoc logging behaviors, which provides qualitative context for understanding the motivations and ad-hoc log-authoring practices. 
(3) We introduced a novel mining method that captures data previously considered inaccessible or hidden in local environments, such as ad-hoc logs, by leveraging developer mistakes.
(4) We collected and published JavaScript, Java, and Python commits from GitHub Archive~\cite{github-archive} and Google Cloud BigQuery~\cite{bigquery}, containing 882,438 removed logs in total, along with corresponding metadata from the GitHub APIs~\cite{github-rest-api-docs}.
\section{Background \& Motivation}
In this section, we examine two areas of related research. First, we review prior work on how developers comprehend the dynamic behavior of software, which motivates our study. Second, we discuss research on production logging, which informs our methodology and shapes our research questions.

\subsection{Comprehending Dynamic Behaviors of Software} Developers must construct accurate mental models of the systems they build, yet this task is often challenging and time-consuming. \cite{gilmore_models_1991} conducted studies of over 80 developers performing debugging tasks and noted that “the success of experts at debugging is not due to better debugging skills, but rather to better comprehension.” Empirical studies report that developers spend between 35\% and 58\% of their time on program comprehension (\cite{ko_exploratory_2006,xia_measuring_2018}). 
Among such comprehension challenges, understanding dynamic behaviors—such as evolving program states or complex execution flows—remains particularly demanding. To gain insights into run-time behavior, developers frequently insert logging statements, rerun the program, or interact with user interfaces (\cite{maalej_comprehension_2014,beller_dichotomy_2018}).
Even with the emergence of artificial intelligence, one of the most time-consuming and challenging parts of the work for developers is to audit and understand the generated code as found in ~\cite{gordon_co-audit_2023, bird_taking_2023}.

Despite its importance, evaluating developers’ comprehension remains a challenging task. Researchers commonly rely on small participant groups in controlled or within-subject studies (\cite{xia_measuring_2018}), distribute surveys that trade qualitative nuance for quantitative breadth (\cite{beller_dichotomy_2018,maalej_comprehension_2014}), or employ proxy measures such as cognitive load or Cyclomatic Complexity (\cite{rilling_identifying_2003}). As noted by \cite{ko_practical_2015}, studies involving human participants demand careful recruitment and management, which constrains scalability even though they provide rich qualitative insights. In contrast, proxies such as cognitive complexity—derived from program structure rather than grounded in empirical studies of developer cognition—scale more easily but may fail to reflect the actual cognitive challenges developers encounter, as argued in \cite{shepperd_critique_1988,lenarduzzi_does_2023}. Motivated by both the difficulty of recruiting human participants and the limitations of complexity metrics, our work explores the use of empirical data—specifically ad-hoc log placements—to approximate developers’ comprehension challenges.

\subsection{Tracking Back in the Forest: Ad-hoc vs. Production Logs}  

To properly situate the reader, we review and contrast the existing research on production logs to our new work on ad-hoc logs.
When navigating the dense forest of a running program, developers leave different kinds of markers to trace their path.  
Production logs are like stones or trail markers—carefully placed and meant to endure, guiding future maintainers along established routes.  

Research on production logs has been extensive, with much of it emphasizing their role in long-term system monitoring and maintenance.  
We summarize this body of work across four perspectives: \textbf{Why}, \textbf{How}, \textbf{Where}, and \textbf{What}.  

\textbf{Why} do developers write logs?: \cite{li_qualitative_2021} surveyed 66 developers and analyzed 223 logging-related issue reports, identifying four primary benefits of logging: troubleshooting, execution tracking, comprehension support, and bookkeeping.  
In the context of embedded software engineering, \cite{yang_interview_2021} interviewed 28 developers and found that production logs are mainly used for problem localization and performance improvement, rather than for reverse engineering or understanding existing systems.  

\textbf{How} do developers log?: \cite{li_which_2017} showed that developers’ choices of logging \emph{levels} are strongly influenced by existing logging statements and surrounding block types.  
\cite{jiang_log-it_2023} also examined challenges in log authoring, identifying issues such as disorganized log output, insufficiently informative visualizations, frequent context loss due to view switching, and trade-offs in crafting and interpreting logs.  

\textbf{Where} do developers place logs in code?: The placement of production logs has also received significant attention.  
\cite{fu_where_2014} conducted a mixed-methods study at Microsoft, combining code analysis with interviews, and identified five categories of log locations.  
Building on this work, \cite{li_where_2020} analyzed seven large open-source Java systems and reported seven categories of log placement and their distributions.  
Both studies laid the foundation for log-location recommendation systems, which have since been refined through learning-based approaches, such as works in \cite{mastropaolo_using_2022, zhu_learning_2015, li_towards_2020}.  

\textbf{What} data do developers log?: The content of logging statements has received comparatively less attention.  
\cite{liu_which_2021} proposed an RNN-based tool to recommend which variables developers should log, though without deeper characterization of these variables.  
\cite{jiang_log-it_2023} observed that developers frequently rely on template strings and labels to maintain continuity between source code and log output.  

In contrast, ad-hoc logs resemble breadcrumbs—dropped quickly, used once, and often discarded when their purpose has been served.  
Research on these ``breadcrumbs'' remains far more limited, likely due to their transient nature.  
\cite{beller_dichotomy_2018} highlighted their significance, showing that most developers continue to rely on simple print statements for debugging.  
This practical reliance has long been echoed by practitioners: Unix contributor \cite{kernighan_unix_1978} remarked that ``the most effective debugging tool is still careful thought, coupled with judiciously placed print statements.''
To date, however, only \cite{jiang_log-it_2023} have explicitly investigated ad-hoc logging through observational studies. Specifically, they examined six novice and six expert JavaScript programmers to inform the development of their tool, Log-It. Their work focuses on ad-hoc logging as a form of live interaction among developers, source code, and terminal output, with the goal of identifying usability challenges when working with logs during debugging sessions.
Our study, in contrast, adopts a more objective and scalable approach by quantitatively analyzing ad-hoc logs in version control systems. Through this analysis, we infer the code locations, structural contexts, and comprehension challenges that lead developers to introduce ad-hoc instrumentation in real-world projects.

Taken together, studies of production logs provide methodological insights and conceptual frameworks that inform our own inquiry into these ephemeral traces, guiding both our research questions and approach.  
\section{Research Questions~\label{section:rq}}
Building on prior research that explores the dimensions of \emph{Where}, \emph{What}, \emph{Why}, and \emph{How}, we define the following research questions to guide the analysis of our collected data.

\begin{description}
    \item[RQ1:] \emph{Where} in the codebase do developers place ad-hoc logs? 
    \item[RQ2:] \emph{What} information do developers record in ad-hoc logs? 
    \item[RQ3:] \emph{Why} do developers rely on ad-hoc logs?
    \item[RQ4:] \emph{How} do developers manipulate ad-hoc logs (e.g., insert, revise, or remove) to achieve their goals? 
\end{description}
\section{Dataset Curation\label{section:datacuration}}

In this section, we describe the curation process for the two datasets that underpin our empirical study: (1) ad-hoc log statements collected from Java (JAVA), Python (PY), and JavaScript/TypeScript (JS) projects, together with associated metadata such as commit information, repository details, and before-and-after code; and (2) live-stream recordings of developers adding ad-hoc logs for debugging or program comprehension. For reproducibility, the full dataset and the crawling scripts are provided in \cite{logs-commits}.

\mysubsection{Collecting Accidentally Pushed Printf Statements\label{subsec:AdhocLogDataCuration}} 
To collect printf statements that were later removed by developers, we first identified the targeted commits from Google BigQuery and GitHub Archive. The process began by extracting all events of type \emph{PushEvent} from GitHub Archive. 
From these events, we extracted the commits whose messages matched the target regular expression, which was derived from qualitative insights and empirical tests as described below. To capture these accidental commits, we needed a generalizable way of identifying commits that involved the removal of ad-hoc logging. Our solution was to reverse engineer a commit-message pattern that would reliably indicate such removals. We began by sampling 2,000 random repositories for each language. Within each repository, we inspected commits to locate those where developers had actually removed print or logging statements from the code. From this filtered set, we focused on the commit messages themselves and used embeddings to cluster them. We then selected 100 representative (centroid) messages and analyzed their common linguistic structure. Based on this analysis, we constructed a regular expression that could capture the recurring patterns found in those messages.
The resulting regular expression thus reflects an empirical generalization of how developers typically describe ad-hoc log removals in commit messages, making it possible to scale our detection across large datasets.

For fetching the specific commits, we relied on GitHub’s event stream rather than the API due to GitHub's API rate limits. 
We collected \emph{PushEvent}s spanning from February 12, 2011 to January 11, 2020. These events sometimes contained commits going back as far as 2007. To extend coverage, we later added a supplementary sample of commits obtained via Google BigQuery, reaching up to October 1, 2024. In this case, the BigQuery SQL query selected only the first commit from each \emph{PushEvent} in the GitHub Archive and applied the same regular expression filters. For JAVA \& PY: We used Google BigQuery to fetch the \emph{PushEvent}s from 2015 to 2022. In this dataset, we selected all commits in each \emph{PushEvent}, with a maximum 20 (as it is capped by the design of GitHub's PushEvent query). 

After retrieving the commits for all languages, we used GitHub APIs to collect the necessary data that is missing from the event metadata, such as actual files and their diffs in the commits, for subsequent analysis.

\mysubsection{Collecting Production Logging Statements\label{subsec:ProductionLogPreparation}}
To further investigate \emph{where} ad-hoc logs are placed in codebases (RQ1), we contrast their locations with the locations of production logs. This comparison helps distinguish patterns that are specific to ad-hoc logging from patterns that may reflect logging practices more generally. Prior studies~\cite{fu_where_2014, li_where_2020} provide useful insights into where developers place production logs. For example,~\cite{li_where_2020} found that production logs in a corporate environment often appear in error-handling code or around logical branches, such as \texttt{if} statements. However, these distributions may not directly apply to our setting, because our ad-hoc log dataset is drawn primarily from personal and open-source projects. 

To support a like-for-like comparison, we collected production logs from the same repositories in which we identified ad-hoc logs. Specifically, we shallow-cloned the repositories from which the filtered ad-hoc logging statements were collected and that remained publicly available. We then scanned the latest version of each repository to identify logging statements primarily intended for production use, such as Log4j\footnote{\url{https://logging.apache.org/log4j/2.x/index.html}} in Java, the \texttt{logging} module\footnote{\url{https://docs.python.org/3/library/logging.html}} in Python, and Winston\footnote{\url{https://github.com/winstonjs/winston}} in JavaScript. For each production log statement, we extracted the same attributes as for the ad-hoc logs, including the surrounding block type and the complexity of the enclosing function. This process yielded 390,191 production logs in Python, 163,033 in JavaScript, and 807,189 in Java.

\mysubsection{Preparing Ground Truth for Comparison\label{subsec:groundtruth}} 
To estimate the complexity and distribution of function types in public code, we curated three repository samples. The first consists of 1,000 JavaScript repositories created between 2023 and 2024, randomly drawn from ~3,000 candidates retrieved via the GitHub Search API. The second and third each include 1,000 repositories, randomly sampled from GitHub Archive and GitHub BigQuery. For Java and Python, we randomly selected 2,000 repositories each from corpora where ad-hoc logs had been removed. Collectively, these samples provide a baseline for evaluating whether the placement of ad-hoc logs reflects substantive factors—such as function complexity, threading, and asynchrony—rather than convention alone. For example, the higher complexity observed in functions from which ad-hoc logs were removed may partly reflect the inherently greater structural complexity of those specific samples, and thus a related need for comprehension of runtime behaviors in such complex code.

\mysubsection{Live-Streamed Videos}
Live streaming is increasingly adopted by developers to increase visibility, support education, and foster community engagement, as noted in~\cite{kokinda_streaming_2023}. Such data enable unobtrusive observation of behaviors in natural settings, helping reduce observer bias (\cite{mahtani_catalogue_2018}). Despite this potential, live streaming remains underexplored in software-engineering research. Progress has been made by curating and labeling live-streamed videos and by developing the \emph{observer-dev.online} platform for software-engineering researchers as created by \cite{online-dev}. From this dataset, we used all 15 curated videos (total duration: 36:01:30) and conducted a thematic analysis, with both the dataset and our qualitative findings made publicly available (\cite{logs-commits}). Since little is known about why and how developers insert log statements, we relied on the segmentation of each video into debugging episodes (e.g., \textit{“Testing the program and reading the outputs”}, \textit{“Interacting with a file of code (Edit)”}) provided in~\cite{alaboudi_what_2023}. These segments allowed us to efficiently identify moments when developers began adding log statements to source code.

\section{Data Analysis~\label{section:dataanalysis}} 
In this section, we describe our data-analysis process for addressing the research questions above.

\subsection{Analyzing Collected Ad-hoc Logs\label{sec:ad-hoc-logs-analysis}}

To address RQ1 and RQ2, we prepared the dataset that contains three programming languages for analysis by conducting the following preprocessing steps: 

\textbf{Filtering:} Some commits contained messages with file changes that are not relevant to logging statements, or added files outside the target programming languages. To address this, we created diff files by comparing the state of files before and after each commit. For JS, we further excluded library and minified files, applying \cite{google-chrome}’s minified file-detection algorithm. We then examined the \emph{before} versions of files to locate original logging statements (e.g., \emph{console.log} or \emph{System.out.println}). When such statements appeared at line numbers corresponding to deleted lines in the diff, we verified this as an ad-hoc log and recorded them as data points, capturing the statement text along with contextual information (e.g., file path, enclosing function, line number, commit sha, and function complexity metrics) in a separate CSV file.

\textbf{Contextualization:} Logging statements alone provide limited insight, so we enriched them with predefined metadata for static analysis. 
This metadata includes the functions and blocks containing each log, visibility of the functions, annotations, parameter types of log statements, and the Cyclomatic Complexity metric, as discussed in \cite{ebert_cyclomatic_2016}, of each enclosed function. The analysis scripts and metadata are available at~\cite{logs-commits}. To detect removed code lines, we inspected diff files for deleted logging statements. Later, we parsed the abstract syntax tree of each targeted source file, traversing from each node to identify its enclosing blocks and functions. As JavaScript allows developers to define anonymous functions, we inferred names for such a function as illustrated in Figure~\ref{fig:customized_function_names}. Our scripts also support analysis of both major Python versions: 3.X (190,695 print statements) and 2.X (29,653 print statements).

\textbf{Validation}: Because our dataset is derived from simply removed logging statements, there is the potential risk that some collected cases may be core-functionality-based logging --- not specifically ad-hoc logs used for runtime comprehension. 
For example, developers may remove print statements that were originally intended for user-facing output or other purposes. 
To account for this possibility, we sought to assess the reliability of two options for log curation: one option that is this looser removal-only scheme, and one option that is stricter, which requires more explicit knowledge of developer intention of some debugging or comprehension activity.
We anticipate a precision--coverage tradeoff between the two: the looser scheme might admit more false positives, whereas the stricter scheme might exclude valid ad-hoc logs.
Specifically, the looser matching scheme identifies commit messages that contain both a removal-related verb and a noun referring to a common debugging statement, such as \texttt{console.log} or \texttt{print}. The stricter one requires a removal-related verb, an adjective indicating debugging or accidental intent, and a noun describing the removed statement. This stricter pattern is intended to capture commit messages such as ``remove accidental print statements'' or ``delete debugging console.log.'' The complete regular expressions are provided in the replication package.
\begin{table}[t]
\caption{Comparison of deleted logging/print statements across languages. \textit{S} denotes the dataset collected with the stricter regex, which requires commit messages to explicitly indicate accidental or debugging intent. \textit{L} denotes the dataset collected with the looser regex, which only requires commit messages to indicate the removal of logging or print statements.\label{tab:deleted-logs}}\centering
\begin{tabular}{|l|r|r|r|}
\hline
\textbf{Language} & \textbf{\# Commits /w Logs} & \textbf{\# \textit{L} Logs} & \textbf{\#\textit{S} Logs}\\
\hline
JavaScript & 215,160 & 548,880 & 15,090\\
Python     & 74,392  & 220,348 & 27,049\\
Java       & 30,093  & 113,210 & 16,975\\
\hline
\end{tabular}
\end{table}

For each language and filtering scheme, we randomly sampled 100 commits from
the commits containing logs collected by that scheme and examined every
removed print or logging statement in the selected commits. The codebook,
provided in the replication package, classified a statement as ad-hoc when its
change and surrounding code indicated temporary debugging or runtime
comprehension use. This included lightweight checkpoint-style messages (e.g.,
\texttt{print("here")}) and direct printing of variables. Statements serving
application behavior, such as output shown to a user through a command-line or
user interface, were not classified as ad-hoc logs. The manually annotated
sampled commits and codebook are available at \cite{manual-commits-label}.

The first two authors independently labeled the samples. They then held a four-hour consensus meeting in which they reviewed all sampled statements, discussed disagreements against the codebook, and assigned the final labels
reported here.

In the looser dataset, manual labeling showed that most sampled logs were
indeed ad-hoc: 264 of 269 in JavaScript (98.14\%), 276 of 319 in Java
(86.52\%), and 148 of 183 in Python (80.87\%). In the
stricter dataset, the proportions were even higher: 156 of 156 in JavaScript
(100\%), 218 of 235 in Java (92.77\%), and 161 of 168 in Python (95.83\%). However,
Table~\ref{tab:deleted-logs} also shows a great difference in dataset
size between the two schemes. The looser scheme collected 548,880
JavaScript logs, whereas the stricter scheme retained 15,090.

Although we sampled 100 commits for each language and filtering scheme, we labeled every removed print or logging statement within those commits.
Therefore, the reported validation results are statement-level proportions. 
Representative false positives included print statements replaced by application or third-party logging facilities and user-facing output removed during broader logic changes.

Because validation began with commits already selected by the regular
expressions, it estimates precision among matched commits but cannot estimate how many valid ad-hoc-log removals occurred in unmatched commits. 
We therefore describe the reduction from the loose to the strict dataset as a coverage tradeoff rather than as a measured recall value. 
Even the stricter filter
should not be interpreted as guaranteeing perfect precision, since developers may use inaccurate or recycled commit messages that do not faithfully describe the actual change.
We report results from datasets in the following sections, denoting the looser dataset as \textit{L} and the stricter dataset as
\textit{S}.

These results also suggest that JavaScript's \texttt{console.log} is a particularly reliable indicator of ad-hoc logging, whereas \texttt{print} statements in Java and Python serve a broader range of purposes, including user-facing command-line output, making them harder to distinguish from non-ad-hoc usage.
Despite this challenge in Java and Python, our above validation found 80--88\% true ad-hoc logs in the looser set and 92--96\% in the stricter set, which should provide an accurate-enough dataset to allow us to derive insights at scale.

\begin{figure}[t]
    \footnotesize
    \begin{align*}
\mathrm{Variable~Declarator} &: \mathtt{var~\textcolor{red}{cb}~=~function~()~=>~\{ \ldots \}} \\
\mathrm{Call~Expression}     &: \mathtt{fetch(\ldots).\textcolor{red}{then}(()~=>~\{ \ldots \})} \\
\mathrm{Assignment~Expression} &: \mathtt{\textcolor{red}{cb}~=~function()~=>~\{\ldots\}} \\
\mathrm{Class~Property} &: \mathtt{class~Text~\{ \textcolor{red}{getText}~=~()~=>~\{\ldots\}~\}}\textbf{}
    \end{align*}
    \caption{Custom approach to assign names to anonymous functions in JavaScript. Tokens highlighted in red represent the names assigned to these functions. We use such method to conduct analysis on function names later.} 
    \label{fig:customized_function_names}
    \vspace{-.1in}
\end{figure}


\begin{figure}
    \centering
    \includegraphics[width=0.7\linewidth]{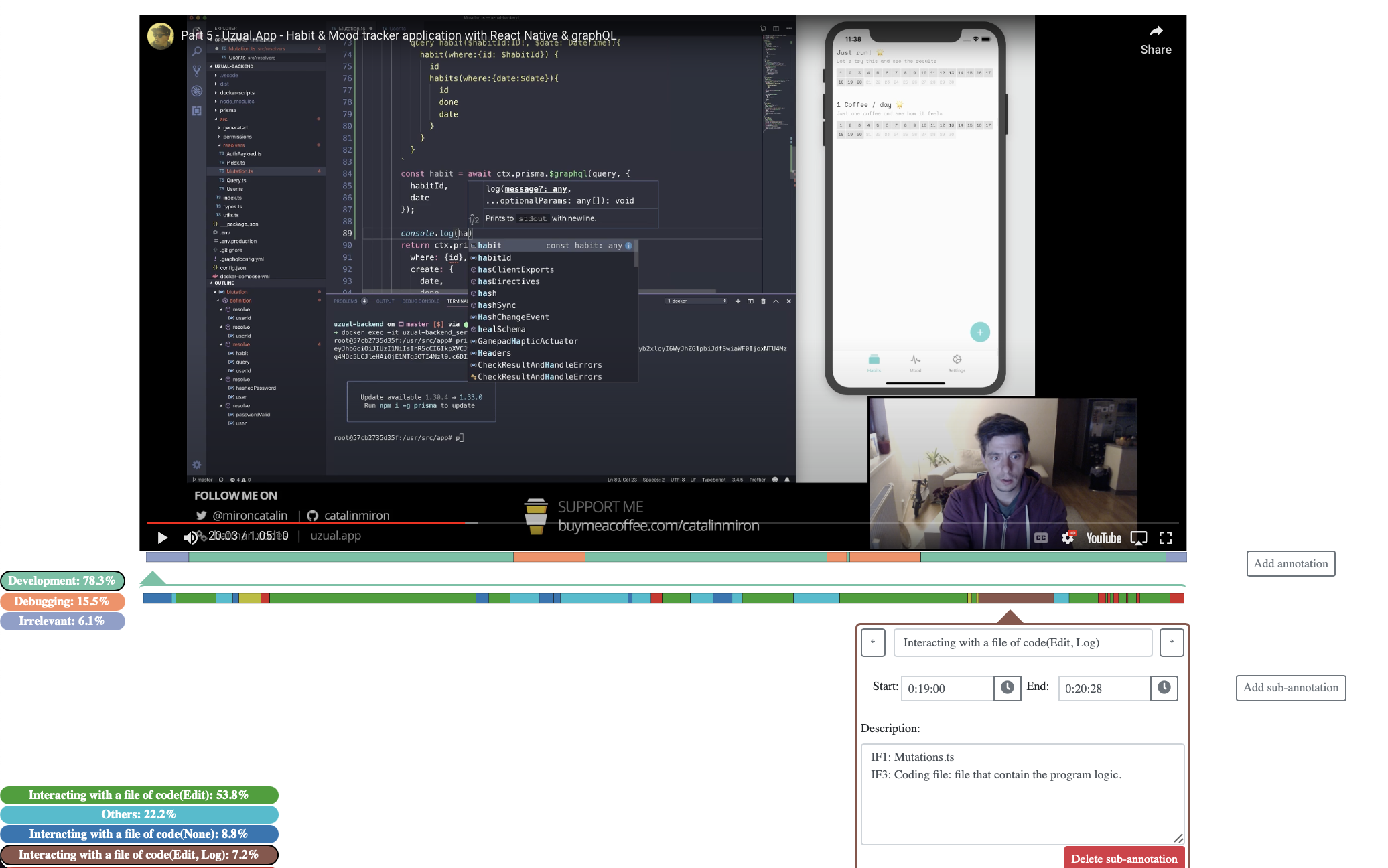}
    \caption{Online YouTube Videos with the Annotation Platform Observe.dev}
    \label{fig:screenshot_of_observer_online}
\end{figure}
\subsection{Qualitative Analysis of Live Streaming Data}
To investigate how developers employ ad-hoc logs for addressing RQ3 and RQ4, we analyzed live-streaming data by first locating debugging episodes within the labeled dataset created by~\cite{alaboudi_what_2023}. We focused on instances marked as “interacting with a file of code (Edit, Log)” or “interacting with a file of code (Log).” For each segment, we replayed the recording—rewinding or fast-forwarding as needed—until we could respond to a set of predefined guiding questions used in our deductive thematic analysis as suggested in~\cite{braun_thematic_2012}.

Figure~\ref{fig:screenshot_of_observer_online} illustrates a screenshot from Observe.dev, capturing a developer inserting a log into their codebase. We used the following guiding questions to examine streamers’ coding actions and verbal explanations: 
\begin{itemize} 
    \item How and why did they insert the log statement?
    \item When and how did they remove the log statements?
    \item Where and what exactly did they log?
\end{itemize}

The initial qualitative coding phase began with the first author reviewing five videos, identifying 45 ad-hoc logs, and developing an initial codebook, which is provided in the replication package at \cite{logs-commits}. The second author then independently coded the same set of videos using this codebook. Inter-rater reliability between the two coders was high (Cohen’s $\kappa = 0.849$). Any disagreements or uncertainties were subsequently resolved through team discussions during weekly meetings until consensus was reached. After this calibration process, first author then proceeded with coding the remaining dataset. In total, fifteen videos from~\cite{alaboudi_what_2023}, amounting to more than 36 hours of material, were analyzed, resulting in the identification of 92 ad-hoc logs. Videos were labeled sequentially from V1 to V15. Three videos contained no ad-hoc logs: V5 and V6 (Rust) relied solely on compile-time error logs, while V14 (CSharp) consistently used the Visual Studio debugger.


\section{Results~\label{sec:result}}

\begin{figure*}[t]
    \centering
    \includegraphics[width=\textwidth]{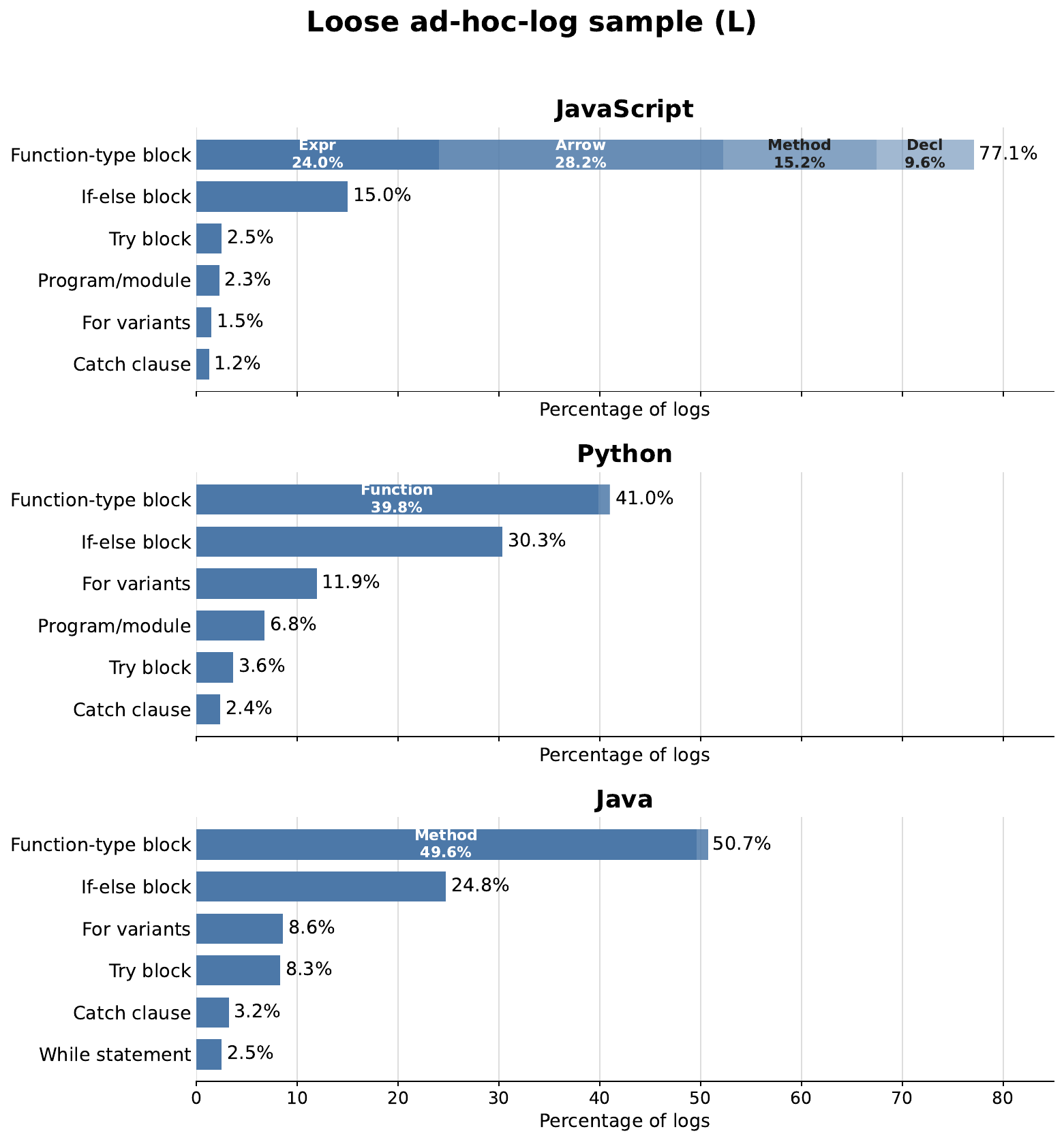}
    \caption{The six most frequent enclosing block types in the looser
    ad-hoc-log sample (L), shown separately for JavaScript, Python, and Java.}
    \label{fig:distribution_of_block_statements}
\end{figure*}

\begin{figure*}[t]
    \centering
    \includegraphics[width=\textwidth]{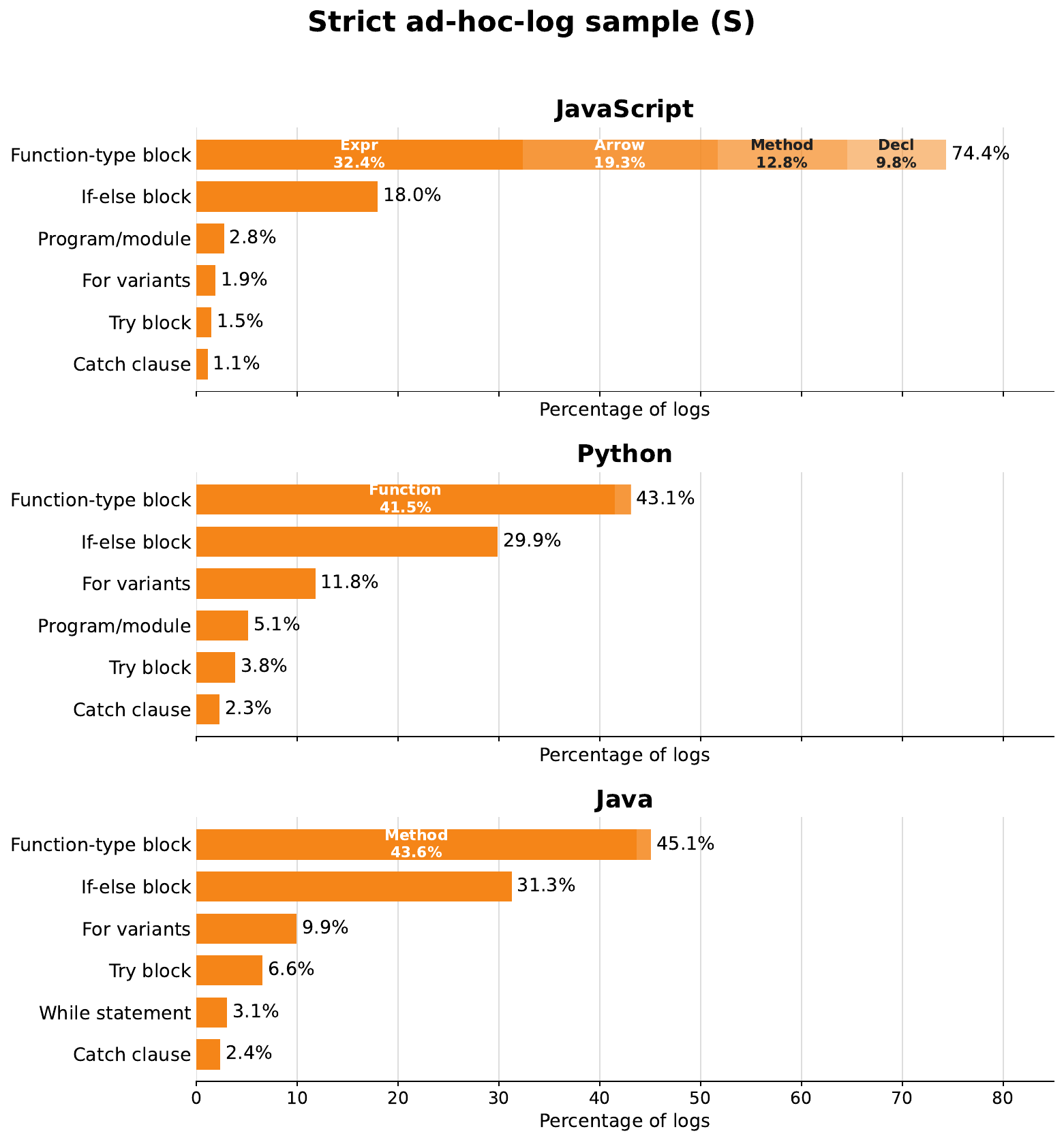}
    \caption{The six most frequent enclosing block types in the stricter
    ad-hoc-log sample (S), shown separately for JavaScript, Python, and Java.}
    \label{fig:distribution_of_block_statements_for_S}
\end{figure*}

\begin{figure*}[t]
    \centering
    \includegraphics[width=\textwidth]{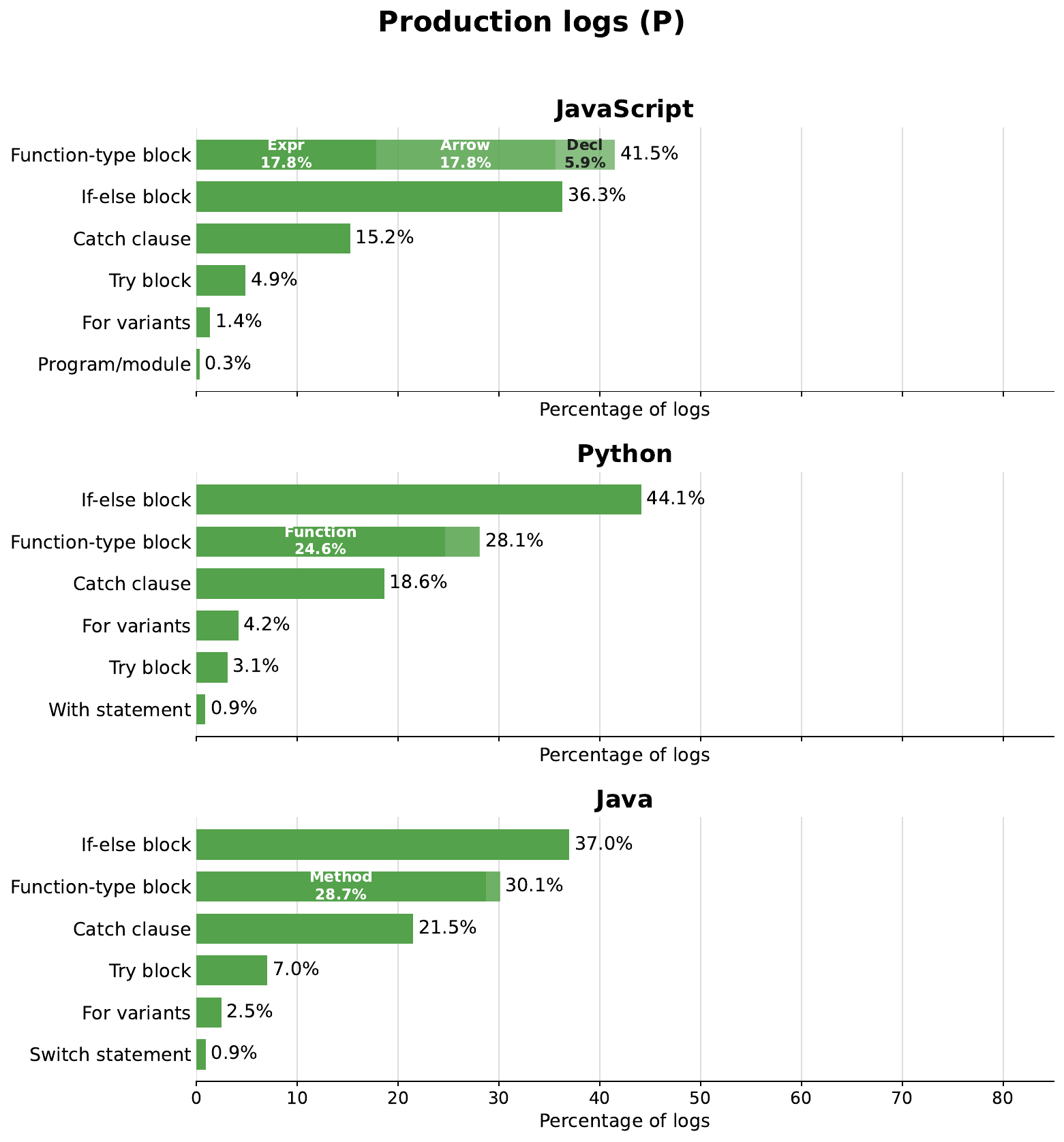}
    \caption{The six most frequent enclosing block types among production logs,
    shown separately for JavaScript, Python, and Java.}
    \label{fig:distribution_of_production_block_statements}
\end{figure*}

Using the methodology described in the prior two sections, we produced a number of results and findings, which we present here, organized by the research questions that we seek to answer.

\subsection{\textbf{RQ1:} Where in the codebase do developers place ad-hoc logs?}
Figures~\ref{fig:distribution_of_block_statements}
and~\ref{fig:distribution_of_block_statements_for_S} show where developers
placed ad-hoc logs across JavaScript, Java, and Python. 
The largest share appears in \textbf{function-type blocks} in all three languages in both stricter and looser datasets: \textbf{77.1\% (L) and 74.4\% (S)} in JavaScript, \textbf{50.7\% (L) and 45.1\% (S)} in Java, and
\textbf{41.0\% (L) and 43.1\% (S)} in Python. By ``function-type block,'' we mean that the log
statement resides directly in the body of a function or method, rather than
nested within a conditional, loop, or error-handling block. This distinction
separates, for example, a \texttt{console.log} at the top of a function from
one placed inside an \texttt{if} branch or a \texttt{try} block. Because the
language datasets differ in collection window, event sampling, extraction and
filtering, and validation precision (Sections~4 and~5.1), their absolute
percentages should not be compared as estimates of language effects. We use
them to characterize placement patterns within each language. 
If-else placements are the next largest group: \textbf{15.0\% (L) and
18.0\% (S)} in JavaScript, \textbf{24.8\% (L) and 31.3\% (S)} in Java, and
\textbf{30.3\% (L) and 29.9\% (S)} in Python. Error-handling blocks are a
minority in the ad-hoc-log samples. For example, JavaScript
\texttt{TryStatement} and \texttt{CatchClause} placements together account
for less than 4\% in both L and S.

These patterns differ from \cite{li_where_2020}'s findings on
\emph{production} logs, which showed that logging statements are commonly
placed in error-handling and conditional blocks. \cite{li_where_2020} focused
on industrial Java systems, whereas our dataset is primarily drawn from
open-source projects across multiple languages. To support a like-for-like
comparison, we also collected production logs from the same repositories using
the procedure described in \S\ref{subsec:ProductionLogPreparation} and
examined their surrounding block types, as shown in
Figure~\ref{fig:distribution_of_production_block_statements}. Our
production-log sample is broadly consistent with prior work: production logs
are more concentrated in \texttt{catch}/\texttt{except}-like blocks and
\texttt{if}--\texttt{else} branches. For example, in our JavaScript
production-log sample, \texttt{CatchClause} accounts for 15.25\% of logging
statements and \texttt{IfStatement} accounts for 36.28\%, compared with 1.2\%
and 15.0\%, respectively, in the JavaScript L sample. Java shows the same
pattern: production logs are concentrated in \texttt{CatchClause} (21.49\%)
and \texttt{IfStmt} (36.98\%), with an additional 7.02\% appearing in
\texttt{TryStmt}. Ad-hoc logs are less frequent in these locations and are
more often placed directly in function or method bodies.

This contrast is visible across languages. In JavaScript, function-related
blocks remain prominent even for production logs: \texttt{FunctionExpression}, \\
\texttt{ArrowFunctionExpression}, \texttt{MethodDefinition}, and
\texttt{FunctionDeclaration} together account for 41.51\% of the sample.
Across all three languages, production logs are more concentrated in
exception-handling and conditional blocks, whereas ad-hoc logs are more
concentrated at the immediate function- or method-body level. Taken together,
this pattern suggests that ad-hoc logs often serve as short-term aids for
runtime comprehension during active development, whereas production logs more
often record branch outcomes and error states in deployed systems.

\vspace{5pt} 
\noindent\fbox{\begin{minipage}{\textwidth}
\textbf{Finding 1A}:
Ad-hoc logs were placed primarily at the top level of function blocks (e.g., 77.1\% in JS, 41.0\% in PY in L set), in contrast to production logs, where about half reside in catch handlers or branch blocks as discussed in \cite{li_where_2020, fu_where_2014, yuan_characterizing_2012}.
\end{minipage}}
\vspace{5pt}

\textbf{Types of Functions with Ad-hoc Logs.} Understanding JavaScript systems is often difficult for developers because of the language’s heavy reliance on asynchronous and event-driven functions. As such, this complexity has motivated significant research on visualization and static analysis techniques aimed at improving debugging for such features (e.g., \cite{alimadadi_understanding_2016, turcotte_drasync_2022}). 


In this work, we investigate whether ad-hoc logs are more commonly used in certain function types—specifically anonymous, callback, and asynchronous functions—to aid developers’ runtime understanding. To establish a baseline, we applied the sampling method described in Section~\ref{sec:ad-hoc-logs-analysis} and report the distributions of function types across three samples in Table~\ref{tab:distribution_of_function_type}. Results show that anonymous functions dominate overall, while asynchronous functions account for less than 10\%, and callbacks comprise fewer than 40\%. 
Async functions are defined with the \emph{async} keyword and typically manage asynchronous tasks such as HTTP requests. We label functions declared inline at a call site as “callback functions.” Anonymous functions, as expected, are those without names.

\begin{table}[t]
    \caption{Function-type distributions in the three JavaScript baseline
    samples. Cells report count (percentage). Categories can overlap, so row
    percentages do not sum to 100\%.\label{tab:distribution_of_function_type}}
    \centering
    \begin{tabular}{@{}llrrr@{}}
        \toprule
        Sample & Function type & Total \#functions & Count & Percentage \\
        \midrule
        \multirow{4}{*}{2023--2024}
          & Async     & \multirow{4}{*}{1,070,838} & 46,940  & 4.38\% \\
          & Anonymous &                            & 664,649 & 62.07\% \\
          & Callback  &                            & 249,743 & 23.32\% \\
          & Other     &                            & 377,939 & 35.29\% \\
        \addlinespace
        \multirow{4}{*}{GitHub Archive}
          & Async     & \multirow{4}{*}{2,937,684} & 18,628    & 0.63\% \\
          & Anonymous &                            & 2,393,852 & 81.49\% \\
          & Callback  &                            & 902,542   & 30.72\% \\
          & Other     &                            & 516,581   & 17.58\% \\
        \addlinespace
        \multirow{4}{*}{Google BigQuery}
          & Async     & \multirow{4}{*}{2,009,471} & 146,351   & 7.28\% \\
          & Anonymous &                            & 1,725,954 & 85.89\% \\
          & Callback  &                            & 771,356   & 38.39\% \\
          & Other     &                            & 262,625   & 13.06\% \\
        \bottomrule
    \end{tabular}
\end{table}

\begin{figure}
    \centering
    \includegraphics[width=1\linewidth]{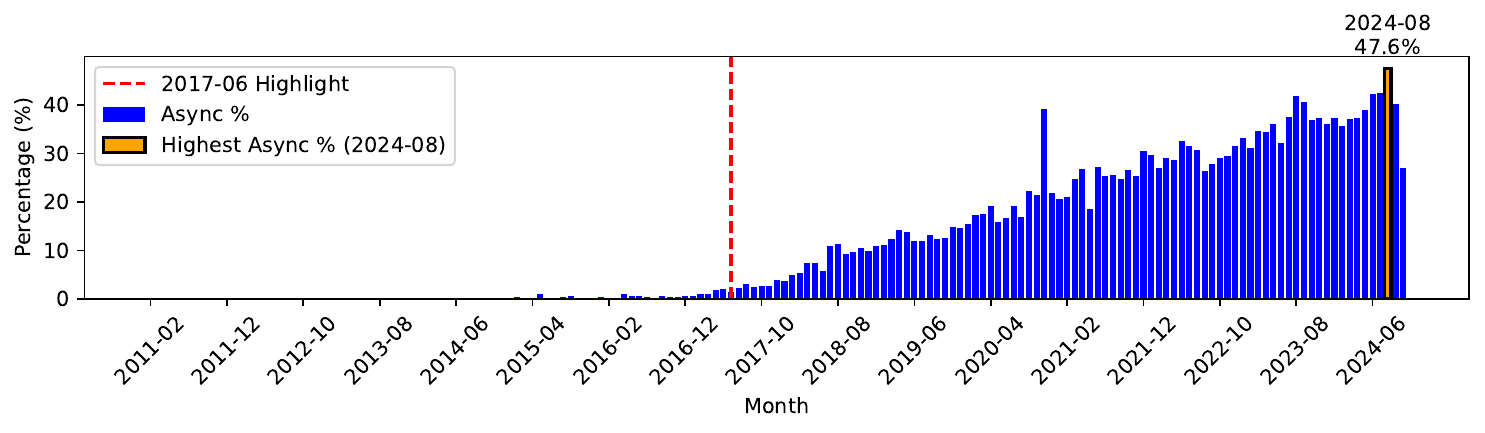}
    \vspace{-.25in}
    \caption{Percentage of ad-hoc logs within asynchronous functions over time in sample \textit{L}~\label{fig:async_func}}
    \label{fig:enter-label}
\end{figure}

We also examined how these distributions changed over time. Figure~\ref{fig:async_func} shows only the trend for asynchronous functions in sample \textit{L}, since the proportions for callback and anonymous functions remain largely stable; the full results for those categories are included in the replication package at ~\cite{logs-commits}. We report only the trend for sample \textit{L} for two reasons. First, manual inspection showed that the JavaScript data has high accuracy in both the looser and stricter samples. Second, the stricter sample contains too few commits and removed logs to reveal a clear temporal trend. Notably, the share of ad-hoc logs appearing in asynchronous functions has risen steadily since the introduction of the \emph{async} keyword in mid-2017, reaching 47.6\% by August 2024. This figure contrasts sharply with our 2023--2024 repository sample, in which only 4\% of all collected functions are asynchronous. Similarly, our qualitative analysis of JavaScript tutorial videos shows that 45.3\% of ad-hoc logs (24 out of 53) appear in async or callback functions.

Prior literature, such as
\cite{alimadadi_understanding_2016,trumper_understanding_2010}, suggests
that multithreading and concurrent programming are particularly challenging
for developers to understand. To examine this in our context, we compared
classes containing removed ad-hoc logs with a baseline drawn from 2,000 Java
repositories.

For each Java source file, the analyzer examined the first top-level type declared in that file. We classified that type as a ``concurrency-type class'' when its declaration directly extended a recognized concurrency-related class
(e.g., \texttt{Thread}) or directly implemented a recognized
concurrency-related interface (e.g., \texttt{Runnable}). 
The detector checks the type names
written in the \texttt{extends} and \texttt{implements} clauses; it does not infer indirect inheritance. 
In the baseline, 3,165 of 760,399 classes (0.42\%) matched the indicator. Among
classes containing at least one removed ad-hoc log statement, 975 of 44,166
class observations (2.21\%) matched it, corresponding to a risk ratio of 5.30.
Thus, concurrency-type classes were more frequent in the log-removal sample,
although they remained a small minority in both groups. These percentages
describe the direct source-code relationships captured by the indicator, not
all forms of Java concurrency.

\begin{table}[t]
\caption{Top-10 method/function names with ad-hoc logs across Java, JavaScript (functions), JavaScript (callbacks), and Python.}
\centering
\begin{tabular}{lllll}
\hline
\textbf{Rank} & \textbf{Java} & \textbf{JS (Functions)} & \textbf{JS (Callbacks)} & \textbf{PY} \\
\hline
1  & test\_0          & render            & then                & main \\
2  & main             & constructor       & on                  & \_\_init\_\_ \\
3  & run              & componentDidMount & for each            & run \\
4  & execute          & success           & catch               & forward \\
5  & test\_alter\_first& handleSubmit     & subscribe           & train \\
6  & actionPerformed  & exports           & get                 & get \\
7  & update           & init              & post                & update \\
8  & handle           & ngOnInit          & map                 & post \\
9  & doGet            & update            & it                  & \_\_call\_\_ \\
10 & initialize       & onSubmit          & add event listener  & \_lint\_html\_files \\
\hline
\end{tabular}

\label{tab:adhoc-hotspots}
\end{table}

\textbf{Name of the functions with ad-hoc logs.}
Across languages, we observed from two perspectives: (i) the names of functions or methods themselves where ad-hoc logs are present, and (ii) in JavaScript specifically, the names of functions that \emph{accept} callbacks containing ad-hoc logs. When considering functions and methods directly, we find that each ecosystem reflects its dominant paradigm. In JavaScript, ad-hoc logs are concentrated in lifecycle methods such as \emph{render}, \emph{constructor}, \emph{componentDidMount}, and \emph{ngOnInit}, pointing to the prevalence of logging in frontend component initialization and update flows. In Java, the most frequent names include \emph{main}, \emph{run}, and \emph{execute}, which correspond to entry points and execution control, but also testing and GUI-related names such as \emph{test\_0}, \emph{test\_alter\_first}, and \emph{actionPerformed}. In Python, the top names are more generic, reflecting its use across diverse domains: common lifecycle functions (\emph{\_\_init\_\_}, \emph{\_\_call\_\_}), general execution (\emph{main}, \emph{run}, \emph{update}), and some workload-specific routines, for example in machine learning (\emph{forward}, \emph{train}) and developer tooling (\emph{\_lint\_html\_files}). Unlike Java and JavaScript, no single pattern dominates beyond the natural diversity of Python applications. Meanwhile, in JavaScript’s callback-accepting functions, the most common names are \emph{then}, \emph{catch}, and \emph{on}, highlighting asynchronous and event-driven control. Taken together, these distributions suggest that ad-hoc logs appear where developers most often need to reason about complex, framework-controlled behaviors: lifecycle timing in JavaScript UIs, concurrency and control flow in Java, and a wide range of tasks shaped by Python’s flexibility across domains.

\vspace{5pt} 
\noindent\fbox{\begin{minipage}{\textwidth}
\textbf{Finding 1B}: In JavaScript, many ad-hoc logs occur in callbacks, asynchronous functions, and library lifecycle functions. The proportion occurring in asynchronous functions increased from the standardization of \smalltt{async}/\smalltt{await} in 2017 through 2024. In Java, concurrency-related classes were more frequent among classes containing removed ad-hoc logs than in the baseline (2.21\% versus 0.42\%). Together, these descriptive results suggest that developers use ad-hoc logs in contexts where indirect or framework-controlled runtime behavior may be difficult to observe.
\end{minipage}}

\vspace{5pt} 
Because the complexity distributions are skewed, we compared the groups using
two-sided Mann--Whitney $U$ tests and report Vargha--Delaney $A$ as the effect
size. The effect size is the probability that a randomly selected observation
from the first group has greater complexity than one from the comparison
group, with ties receiving half weight; $A=0.5$ indicates no distributional
tendency.
\begin{table}[h]
    \centering
    \caption{Cyclomatic complexity of all functions and functions containing
    removed ad-hoc logs. The Mann--Whitney $p$-values and Vargha--Delaney
    $A$ effect sizes compare each ad-hoc group with all functions.}
    \label{tab:cyclomatic-stats}
    \begin{tabular}{llrrrrr}
        \toprule
        Dataset & Group & Mean & Median & Std. & $p$ vs. All & $A$ vs. All \\
        \midrule
        \multirow{3}{*}{BigQuery (JS)}
          & All        & 3.32 & 2.0 & 6.20   & --       & --    \\
          & Ad-hoc (L) & 7.77 & 4.0 & 19.19  & $<0.001$ & 0.699 \\
          & Ad-hoc (S) & 7.80 & 4.0 & 11.96  & $<0.001$ & 0.711 \\
        \addlinespace
        \multirow{3}{*}{GitHub (JS)}
          & All        & 3.64 & 2.0 & 155.06 & --       & --    \\
          & Ad-hoc (L) & 5.62 & 3.0 & 10.21  & $<0.001$ & 0.616 \\
          & Ad-hoc (S) & 6.91 & 3.0 & 14.37  & $<0.001$ & 0.652 \\
        \addlinespace
        \multirow{3}{*}{Java}
          & All        & 1.88 & 1.0 & 3.43  & --       & --    \\
          & Ad-hoc (L) & 6.55 & 3.0 & 12.51 & $<0.001$ & 0.762 \\
          & Ad-hoc (S) & 7.86 & 4.0 & 14.16 & $<0.001$ & 0.824 \\
        \addlinespace
        \multirow{3}{*}{Python}
          & All        & 2.56 & 1.0 & 4.14  & --       & --    \\
          & Ad-hoc (L) & 6.99 & 3.0 & 23.56 & $<0.001$ & 0.597 \\
          & Ad-hoc (S) & 7.32 & 4.0 & 12.41 & $<0.001$ & 0.691 \\
        \bottomrule
    \end{tabular}
\end{table}

\begin{table}[h]
    \centering
    \caption{Cyclomatic complexity of functions containing production logs
    and rank-based comparisons with the ad-hoc-log datasets. Vargha--Delaney
    $A$ is oriented as production versus ad-hoc.}
    \label{tab:production-log-complexity}
    \begin{tabular}{lrrrcccc}
        \toprule
        Language & Mean & Median & Std. & $p_L$ & $A_{P,L}$ & $p_S$ &
        $A_{P,S}$ \\
        \midrule
        JavaScript & 8.59  & 5.00 & 13.69 & $<0.001$ & 0.590 &
                     $<0.001$ & 0.577 \\
        Java       & 9.09  & 4.00 & 29.18 & $<0.001$ & 0.572 &
                     $0.007$ & 0.506 \\
        Python     & 11.38 & 6.00 & 36.79 & $<0.001$ & 0.666 &
                     $<0.001$ & 0.619 \\
        \bottomrule
    \end{tabular}
\end{table}

\begin{figure}
    \centering
    \includegraphics[width=1\linewidth]{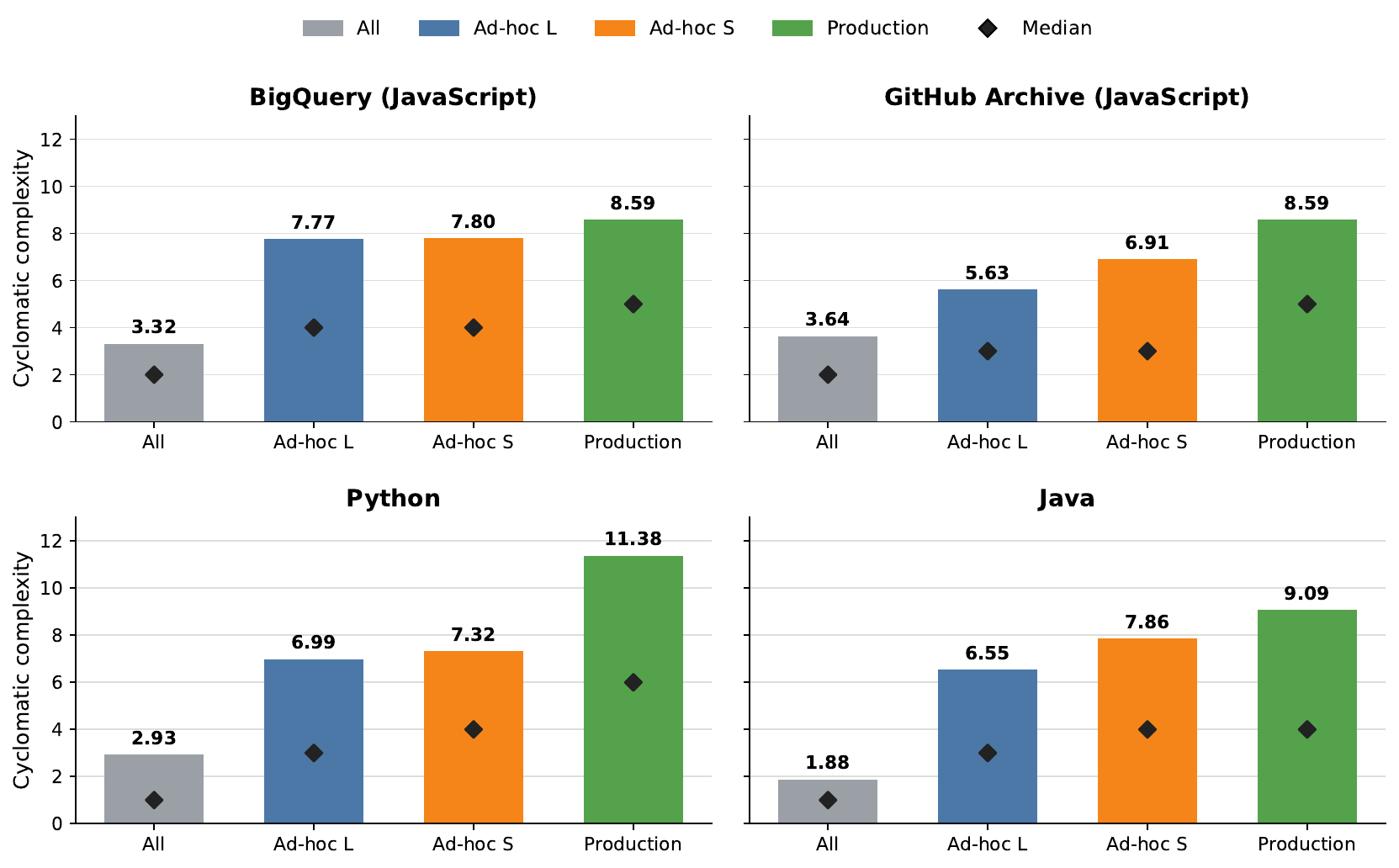}
\caption{Cyclomatic complexity of all functions, functions with removed ad-hoc logs, and functions with production logs across datasets. Bars show mean complexity, while diamond markers indicate medians.}    \label{fig:cyclomatic-graph}
\end{figure}

We examined whether functions containing removed ad-hoc logs differed in observed cyclomatic complexity from the sampled function populations. Comparing cyclomatic complexity across datasets, we found that logged functions are consistently more complex. In JavaScript (BigQuery and GitHub Archive), logged functions had higher mean and median complexity than all functions (BigQuery: $\mu=7.77$, $\tilde{x}=4.0$ vs. $\mu=3.32$, $\tilde{x}=2.0$; GitHub: $\mu=5.62$, $\tilde{x}=3.0$ vs. $\mu=3.64$, $\tilde{x}=2.0$), with $p<0.001$ in both cases. The same pattern appeared in Java ($6.55$ vs. $1.88$, $p<0.001$) and Python ($6.99$ vs. $2.56$, $p<0.001$).   Taken together, functions containing removed ad-hoc logs had higher observed
cyclomatic complexity across the studied datasets, as summarized in Table~\ref{tab:cyclomatic-stats} and Figure~\ref{fig:cyclomatic-graph}.

Functions containing production logs also had higher mean complexity than functions containing removed ad-hoc logs in all three languages (JavaScript: 8.59; Java: 9.09; Python: 11.38). The rank-based effect sizes, however, show that the magnitude varies across comparisons. For production logs versus \textit{L}, Vargha--Delaney $A$ was 0.590 in JavaScript, 0.572 in Java, and 0.666 in Python. Against \textit{S}, it was 0.577, 0.506, and 0.619, respectively. Thus, the Java production-versus-\textit{S} distributions were nearly indistinguishable despite the difference in their means. These results are summarized in Table~\ref{tab:cyclomatic-stats}, Table~\ref{tab:production-log-complexity}, and Figure~\ref{fig:cyclomatic-graph}. These comparisons describe unadjusted associations; function size, project characteristics, and other structural properties may influence both cyclomatic complexity and logging activity.

\vspace{5pt} 
\noindent\fbox{\begin{minipage}{\textwidth}
\textbf{Finding 1C}: Across the studied datasets, functions containing removed ad-hoc logs had higher observed cyclomatic complexity than the sampled overall
function populations.
\end{minipage}}
\vspace{5pt} 
\subsection{\textbf{RQ2:} What information do developers record in ad-hoc logs?}

\textbf{Ad-hoc Logs with Variable Labels.}
Most logs contain only a single variable or literal (72.6\% in JS, 80.5\% in PY, 95.1\% in JAVA) or two arguments (24.2\% in JS, 14.0\% in PY, 0.6\% in JAVA). Consistent with prior findings (\cite{jiang_log-it_2023}), developers often add labels to help trace logged variables back to their origins in the console output. Our analysis shows that when a log includes two arguments, 84.7\% in JS and 76.8\% in PY contain at least one literal. (We excluded Java from this analysis since only 0.6\% of print statements have two arguments.) Among these, 51.5\% (JS) and 38.8\% (PY) explicitly reference the other argument within the literal. For example, in \texttt{console.log(\textquotesingle{}Results: \textquotesingle{}, results)}, the string label ``\texttt{Results: }'' clarifies the meaning of the logged variable value, making it easier to locate in the output stream.

\textbf{Ad-hoc Logs Formats.}
Table~\ref{tab:literal-args} presents the top 10 most frequent literal contents in ad-hoc logs for each programming language. The symbol \emph{R} denotes a sequence of characters repeated more than three times; for example, \emph{=R} represents repeated equal signs such as ``\texttt{====},'' often used to create separation markers or to highlight certain content in the log output. The most dominant entries across the three languages are template-like strings, which often serve as lightweight markers in logs. Strings such as \emph{here}, \emph{clicked}, and \emph{hi} are frequently used to indicate control-flow checkpoints or event triggers during execution. In the case of Java, transaction-related lifecycle markers (e.g., \emph{Transaction started}, \emph{Transaction committed}) are prevalent, reinforcing our observation in Finding~1B.

In live-stream video coding, the percentages are similar for the log parameter count, where the logs contain one (69.6\%) or two (19.6\%) arguments, and 15 out of 18 (83.3\%) two-argument logs are structured as a label string and variables. Furthermore, 9 out of 15 (77.8\%) contain the name of the variable. In addition, similar to the JSON.stringify function, developers using C/C++ tend to create formatting functions of their own. In V4, V7, V12, the developers wrote their own debug function \texttt{infof}, \texttt{dbg}, \texttt{dbgprintf} to simplify forming a more readable log output. When logging variables without native serialization methods, the developers in V4 and V7 wrote helper functions to output variables in a readable string.

\begin{table}[t]
\centering
\caption{Top-10 Most Frequent Literal Argument Strings Across Languages}
\label{tab:literal-args}
\begin{tabular}{clll}
\hline
\textbf{Rank} & \textbf{JavaScript} & \textbf{Python} & \textbf{Java} \\
\hline
1  & NUMBER              & -R                 & -R \\
2  & here                & \textbackslash n   & Transaction started .R \\
3  & -R                  & =R                 & NUMBER \\
4  & test                & NUMBER             & Transaction committed    .R \\
5  & hello               & )                  & *R \\
6  & hi                  & here               & =R \\
7  & success             & \#R                & \textbackslash n \\
8  & error               & *R                 & Starting standalone server.R \\
9  & =R                  & '                  & testRead() .R\textbackslash n \\
10 & clicked             & +R                 & Transaction interrupted .R \\
\hline
\end{tabular}
\end{table}

\vspace{5pt}
\noindent\fbox{\begin{minipage}{\textwidth}
\textbf{Finding 2}: To locate variables and to understand execution order, developers structure the output of ad-hoc logs through labeling and formatting strings akin to findings from prior work (\cite{jiang_log-it_2023}).
\end{minipage}}
\vspace{5pt}

\subsection{\textbf{RQ3:} Why do developers rely on ad-hoc logs?}
To answer RQ3 and RQ4, we shifted our focus to live-streamed data to address more qualitative questions and gain insight into why and how developers use ad-hoc logs. From analyzing and coding 15 live-stream videos and 92 obtained ad-hoc logs, we summarize our key findings as follows:

\textbf{Understanding Program States (n=68)}: Developers often use ad-hoc logs to observe or compare program states for several reasons. A common reason is to capture states that are not immediately clear from the source code, such as states introduced by a third-party library or fetched via a web API. For example, the developer in V3 expressed uncertainty about a library function call's effect and logged a boolean value triggered by a GUI click event, noting, ``\textit{Does it return true when option changes, or does it return true when it set it on? That's the thing I don't know. Let's find it out.}'' Developers also log multiple states to cross-reference their values, ensuring they align with expectations. For instance, the developer in V13 checked three boolean values with logs to iteratively verify they matched expectations. In other cases, logs are used to check the state of a single variable or condition, providing quick feedback on specific program values. 
To understand the initial program states when debugging the function, developers also inserted the logs right after the function to inspect the parameters. For example, the developer in V9 logged the parameter right after the function to check how the function works, \textit{``alright, let's see what this does.''}
Developers also crafted their own formatting function when dealing with complex or non-readable program states. For example, the developer in V4 working on the curl project introduced a function \texttt{what2name} to transform a constant integer to a string, which increased the readability of logging outputs.

\textbf{Understanding Execution Flow (n=33)}: Logging is often used to understand or verify code execution, particularly for those in asynchronous functions or executions from third-party libraries. For example, V13 inserted an ad-hoc log with ``{-}{-}{-}{-}{-}'' in a lifecycle method, noting, ``\textit{It gives me a spot to look at... yep, looks like the componentDidUpdate did not get called when I set the props.}'' Another reason is sanity checking (n=31), where developers insert strings or variables to confirm that the program reaches specific points. For example, V1 logged “IM GOING TO RUN HERE” to verify if the software actually reaches to that specific point.

\vspace{5pt}
\noindent\fbox{\begin{minipage}{\textwidth}
\textbf{Finding 3}:  Developers added ad-hoc logs to gain insight into program states and execution flow, particularly in cases that were not immediately obvious—such as asynchronous functions or behaviors introduced by third-party libraries. These logs were also used to validate assumptions about expected runtime behavior and to carry out sanity checks.
\end{minipage}}
\vspace{5pt}

\subsection{\textbf{RQ4:} How do developers manipulate ad-hoc logs (e.g., insert, revise, or remove) to achieve their goals?}
Our study found that developers employed multiple approaches to create and refine ad-hoc logs, such as iterative modification, comparative checks, varied timing, and strategies for removing logs. The main themes derived are:

\textbf{Iterations (n=27):} Developers progressively refined their logs to better understand program states and identify optimal logging points or variables. For instance, V15 revised the same log three times, evolving from \emph{console.log(this.props)} to \emph{console.log(this.props.navigation.action)} across multiple edit–run cycles. Likewise, V7 made six adjustments while tracing the call stack of a double-click event, moving the log back one level each time, and explained, \emph{``I don’t remember how all this works, so we are just gonna print to make sense of it.''}

\textbf{Comparison (n=22)}: Developers leveraged logs to contrast program states, either by grouping several states within one log statement or by distributing them across multiple logs. They often added labels (e.g., variable names) to link outputs back to the source code. For instance, V4 recorded the same variable at two execution points with distinct labels to verify whether the values matched.

Developers also inserted ad-hoc logs at multiple points in the code to compare the sequence of execution, especially within complex logic or asynchronous flows. For example, V2 placed two dummy strings in different locations and tracked their order of appearance in the log output as various events unfolded.

\textbf{Varied Timing of Ad-hoc Log Removal}: As anticipated, most ad-hoc logs were discarded once they fulfilled their immediate purpose. Specifically, many logs were deleted right after their output was inspected (n=61), and most were observed or confirmed to have been eventually removed (n=80). For the remaining cases, some logs could not be confirmed because the recordings ended before their fate was observed (n=7), while others were ultimately committed to the repository (n=5). Among the committed logs, three were intentionally retained as they evolved into production code (n=3), and two appeared to be accidental commits (n=2). Noteworthy instances include V1 storing logs in a \emph{git stash} for potential reuse during future debugging, and V13 being prompted to remove a log immediately after viewing its output in the terminal.

\textbf{Varied Strategies for Ad-hoc Log Removal}: The strategies employed for ad-hoc log removal were diverse. While direct deletion was the predominant approach (n=54), several alternative practices were observed. For example, V1 avoided committing files containing logs (n=2), and V9 temporarily commented out one log to facilitate run-to-run comparisons (n=2). In V13’s case, a linter tool prompted the removal of ad-hoc logs (n=1), and the developer also created and used an external environment (CodeSandbox) to capture debugging outputs in real time without altering the main codebase (n=3). However, not all log management actions were deliberate—V15 inadvertently committed two ad-hoc logs after forgetting to delete them (n=2).

\vspace{5pt}
\noindent\fbox{\begin{minipage}{\textwidth}
\textbf{Finding 4}: We identified two primary strategies in the way developers authored ad-hoc logs: (1) iteratively refining the content or placement of logs, and (2) inserting multiple logs at once to capture information in parallel. The differing timings and varied approaches to removing these logs indicate that their deletion is often as ad-hoc as their creation.
\end{minipage}}
\vspace{5pt}

\section{Discussion~\label{section:discussion}}
\mysubsection{Perils and Promises of Learning from Mistakes}
As Henry Lieberman noted in (\cite{lieberman_debugging_1997}), ``debugging is the dirty little secret of computer science.'' This ``dirt'' has only accumulated over time, with using debuggers to understand runtime behavior often being regarded as the default solution, leading some developers to feel ambivalent or even reluctant to admit their use of ``printf debugging'' techniques (\cite{beller_dichotomy_2018}). Such hesitation limits the research community’s insight into debugging behaviors, like ad-hoc print statement usage, which play a significant role in shaping final patches or feature development. 
Our research uncovers this ``dirt,'' through mining developers' mistakes to offer the community new possibilities to understand ad-hoc logging practices.

\cite{alaboudi_edit_2021} conducted analysis of live-streaming data and highlighted that software development is an iterative process involving frequent edit-run cycles. However, these small cycles are often hidden within local development environments, limiting the feasibility of large-scale studies. Previous research has addressed this challenge by either conducting costly lab studies (\cite{gilmore_models_1991}), analyzing live-streaming videos (\cite{alaboudi_exploratory_2019}), or requiring developers to install intrusive IDE plugins (\cite{aniche_how_2022, beller_dichotomy_2018}). Our method opens new possibilities for scalable insights into these cycles.

While our collection of ad-hoc logs from revision-control systems enables large-scale analysis, there are limitations: (1) the mined data provides only a snapshot of logs unintentionally committed to version control, making it difficult to observe iterations in the log-authoring process; (2) although the surrounding source code offers some context, it does not capture the developers' rationale for adding, removing, or evaluating the utility of these logs. To gain deeper insight into the context, reasoning, creation, and deletion of these logs, we mined and analyzed the live-streaming data.

Moreover, the distribution of the repository reveals that our dataset, similar to the repositories reported in \cite{kalliamvakou_promises_2014}, is skewed toward smaller, personal, and often inactive repositories. We chose not to explicitly exclude personal projects, as this distribution presents a double-edged sword for our analysis. Whereas large, well-maintained open-source projects often enforce linter checks on pull requests to prevent ad-hoc logs, as suggested by \cite{linter}, personal projects---wherein maintainers might carelessly push ad-hoc logs---allowed us to perform large-scale analysis. Future researchers using our dataset or similar methods should proceed with caution, as data without fine-grained filtering may not fully generalize to expert developers or contributors in collaborative open-source projects.

\mysubsection{Markers in the Forest of Complex Software Execution} Each logging statement placed by developers can be understood as a marker left within the forest of execution, intended to make sense of code that would otherwise be difficult to navigate. Our analysis shows that ad-hoc logs are frequently inserted into threading-related code, asynchronous functions, callbacks, and the lifecycle methods of third-party libraries (Finding 1B). These contexts complicate reasoning because the order and timing of execution are often outside the developer’s immediate control. Logs in such locations serve as reference points for reconstructing runtime behavior. In addition, functions containing removed ad-hoc logs had higher observed cyclomatic complexity than the sampled function populations across all three languages (Finding 1C). This association is consistent with developers using logs in code that is structurally difficult to reason about, although the similar pattern among production logs indicates that it is not unique to ad-hoc logging.

While these findings may appear intuitive, the patterns of where developers place logs provide an opportunity to re-examine how software complexity is measured. The debate over which metrics—Cyclomatic Complexity, Cognitive Complexity, or custom alternatives—best capture developer-perceived difficulty has persisted for decades, such as discussion in \cite{shepperd_critique_1988, scalabrino_automatically_2017}. Since controlled human studies remain expensive and difficult to scale, our
dataset offers a complementary empirical lens and motivates future work examining whether logging behavior can complement established measures of code complexity and developer-perceived difficulty. This opens the door for future work to validate or refine existing metrics, or to propose new ones grounded in observed developer practices. In applied contexts, educators and maintainers could also use patterns of ad-hoc logging as a proxy for identifying code that is unusually challenging, helping to target refactoring efforts or calibrate the level of instructional material. For instance, the high frequency of logs used to understand React component lifecycles in JavaScript may point to aspects of the API design that warrant reconsideration or a separate visualization to support comprehension. 

A single marker is rarely enough to navigate complex execution paths. As shown in Finding 4, developers often placed logs iteratively, refining their placement until the information was sufficient to understand program states and flows. This iterative pattern indicates that initial logs are frequently suboptimal for comprehension, functioning more as tentative probes than definitive solutions. These observations highlight an opportunity for future research: by building on insights from production log studies, it may be possible to recommend more effective initial logging locations such as techniques introduced in \cite{fu_where_2014, li_where_2020} and variables like \cite{liu_which_2021}. Such guidance could reduce the trial-and-error effort developers currently expend, allowing them to converge on useful runtime insights more quickly.

Effective logging is not only about where statements are placed, but also about the richness of the information they convey. Finding 2 shows that many ad-hoc logs extend beyond a simple message: developers often include labels, formatting strings, or custom debug helpers. Finding 3 further shows that developers craft anchor strings and selectively structure complex objects so that logs can be compared across states and used to trace execution over time. Together, these findings suggest that developers treat ad-hoc logs as deliberate reasoning aids rather than arbitrary print statements. This observation has direct implications for tool design. IDE support should not only recommend where to insert logs, but also help developers generate informative logs by suggesting discriminative anchor strings, relevant variable groupings, and readable object summaries tailored to the local context. Such support could be especially valuable in asynchronous or callback-heavy code, where developers must mentally map console output back to source locations and execution order. Our findings also suggest practical logging guidelines: developers benefit from logs that include stable textual anchors, explicit state summaries, and enough contextual detail to distinguish nearby execution points. Existing tools such as Log-it~\cite{jiang_log-it_2023} begin to reduce the burden of connecting output to code, while prior work on visualization~\cite{turcotte_drasync_2022} and logging recommendation~\cite{li_towards_2020} points to promising directions for operationalizing these insights in future developer-support tools.

\mysubsection{Exploratory recursion sensitivity analysis} We also considered recursion as a potential source of comprehension difficulty, since prior work suggests that recursive structures and code constructs may introduce cognitive complexity not fully captured by standard cyclomatic complexity as discussed by ~\cite{hao_accuracy_2023,ajami_syntax_2019}.  As a post hoc analysis,
we examined whether direct recursion might explain the complexity pattern in
Finding~1C. This analysis is exploratory and is not part of the confirmatory
RQ1 findings. We applied a source-code self-call indicator consistently to the
baseline and ad-hoc populations within Python and Java.

In Python, the indicator marks a function when its body contains a direct call
written as \texttt{f()} or \texttt{self.f()}, where \texttt{f} is the enclosing
function's name. It identified 792 of 74,219 functions containing removed
ad-hoc logs (1.067\%) and 20,268 of 2,293,489 baseline functions (0.884\%), a
prevalence ratio of 1.208 (95\% CI [1.125, 1.296]). In the retained subset used
for the exclusion sensitivity analysis, removing functions marked as recursive
changed the estimated difference in the geometric mean of one plus cyclomatic
complexity only slightly, from 32.65\% to 32.45\%.

In Java, the indicator marks a method when its body contains a direct call
written as \texttt{f()}, \texttt{this.f()}, or \texttt{C.f()}, where
\texttt{f} is the enclosing method's name and \texttt{C} is the enclosing
class's name. It identified 1,110 of 60,932 methods containing removed
ad-hoc logs (1.822\%) and 316,235 of 12,546,753 baseline methods (2.520\%),
corresponding to a prevalence ratio of 0.723 (95\% CI [0.682, 0.766]).
After methods marked as recursive were excluded, mean cyclomatic complexity
remained higher among methods containing removed ad-hoc logs than among
baseline methods (4.82 versus 1.87).

\mysubsection{Threats to Validity} 

\mysubsubsection{Internal Threats to Validity} For the mined ad-hoc logs, a key internal threat is that some logs classified as ad-hoc may actually be removed production logs, which could affect the accuracy of our findings. To address this, we manually labeled 100 sample commits for each language and found that only a small portion were not used in an ad-hoc manner. In coding the live-streaming data, researcher bias is a common concern due to the subjective nature of qualitative analysis. We mitigated this by involving a second researcher and holding iterative discussions to minimize uncertainty and resolve any disagreements.

Our Java concurrency-type indicator captures the common case in which a method belongs to the primary class in its source file, but it may miss or misattribute methods in nested, anonymous, and secondary types. Because it examines directly declared supertypes, it also does not capture indirect inheritance, executor submissions, lambdas, annotations, or framework and event-dispatch callbacks. The Java comparison therefore concerns this source-code indicator rather than all forms of Java concurrency.

\mysubsubsection{External Threats to Validity}
Our findings may not generalize to other programming languages or development contexts. Although our repository analysis covers Java, JavaScript, and Python, the qualitative analysis includes live-stream videos spanning a broader set of languages, and the two datasets should therefore not be interpreted as directly comparable language samples. The live-stream analysis is also based on a relatively small sample and may be subject to performance bias, as developers may behave differently while streaming their work as noted in both \cite{chou2026building} and \cite{alaboudi_exploratory_2019}. Moreover, the repositories represented in these videos are skewed toward personal or inactive projects, which may not reflect development practices in actively maintained or industrial projects. While the selection and representativeness of these videos were discussed by \cite{alaboudi_exploratory_2019}, our study uses them to investigate different research questions and therefore inherits additional limitations specific to our analysis. These factors limit the generalizability of our qualitative observations, which should be interpreted as exploratory evidence of how developers use ad-hoc logs rather than as representative estimates of broader developer behavior.

\section{Conclusion}

This paper presented the first large-scale empirical study of ad-hoc, or ``printf-style,'' logging by systematically mining accidental commits and analyzing live-streamed programming sessions. 
By leveraging over 800,000 removed log statements across JavaScript, Java, and Python, as well as 36 hours of qualitative observations, we demonstrated that ad-hoc logs are not arbitrary artifacts but purposeful markers developers employ to navigate complex code and runtime behaviors.

Our findings highlight three themes. 
First, ad-hoc logs are disproportionately concentrated in contexts where reasoning about execution is especially difficult, such as asynchronous and callback functions in JavaScript or thread-related code in Java. 
Second, both ad-hoc and production logs tend to occur in functions with higher cyclomatic complexity. This shared pattern is consistent with logging serving as a means of observing runtime behavior in structurally complex functions, whether the logs address an immediate debugging need or remain for long-term production use.
Third, developers refine ad-hoc logs iteratively, structure them with labels and formatting for clarity, and frequently remove them once their immediate purpose is served---though often with ad-hoc strategies for removal.

Taken together, these results contribute both methodological and practical insights. 
Methodologically, our approach demonstrates how accidental artifacts in version control can serve as valuable data sources for understanding hidden aspects of software practice. 
Practically, our results suggest opportunities for tools that recommend effective logging locations, improve log readability, or visualize execution traces, thereby reducing the trial-and-error burden of printf debugging.

Ultimately, ad-hoc logs, though ephemeral, offer a window into developers' struggles and strategies for program comprehension through aggregation. 
By learning from these transient traces, researchers and tool builders can better support the everyday practices that shape debugging and development, bridging the gap between lightweight, local strategies and more formalized approaches to runtime understanding.


\section{Declarations}

\subsection{Funding}
This research is partially supported by the Dieter Schwarz Stiftung Foundation. Yi-Hung Chou is supported in part by the Taiwan Government Scholarship to Study Abroad.
\subsection{Ethical Approval}
Not applicable.

\subsection{Informed Consent}
Not applicable.

\subsection{Author Contributions}
Yi-Hung Chou led the study, designed the experiments, wrote the manuscript, and developed the crawling and analysis scripts used across all languages. Boyuan Jiang conducted the experiments on Java. Vidit Jain conducted the experiments on Python. Yiyang Min performed the qualitative analysis and supported the JavaScript and TypeScript analyses. April Yi Wang and James A. Jones supervised the research, provided guidance on the experimental design, and contributed to manuscript writing and revision.

\subsection{Data Availability Statement}
All datasets generated and analyzed during this study are openly available on \textit{figshare}. The manual check of collected logs through 100 commits sample across all programming languages is available at \url{https://doi.org/10.6084/m9.figshare.27325635.v3}. The replication package, including logs, commits, crawling and analysis scripts, and qualitative analysis on live streams, is available at \url{https://doi.org/10.6084/m9.figshare.27638058}. The additional dataset and scripts for Java, Python, Stricter JS, as well as the manual validation results can be found at \url{https://doi.org/10.6084/m9.figshare.33994521}

\subsection{Conflict of Interest}
The authors declare that they have no conflict of interest.

\subsection{Clinical Trial Number}
Clinical trial number: not applicable.

\begin{acknowledgements}
We extend our gratitude to the anonymous reviewers for their thorough and insightful feedback. We would also like to thank Heng Du, Junling Wang, Xiaotian Su, Zeyu Xiong, Gustavo Umbelino, Behnood Masoudi, Monil Narang for providing feedback and support.

\end{acknowledgements}


\bibliography{references, data}
\bibliographystyle{spbasic}    


\end{document}